
\magnification = 1200
\def\lapp{\hbox{$ {
\lower.40ex\hbox{$<$}
\atop \raise.20ex\hbox{$\sim$}
}
$}  }
\def\rapp{\hbox{$ {
\lower.40ex\hbox{$>$}
\atop \raise.20ex\hbox{$\sim$}
}
$}  }
\def\barre#1{{\not\mathrel #1}}
\def\krig#1{\vbox{\ialign{\hfil##\hfil\crcr
$\raise0.3pt\hbox{$\scriptstyle \circ$}$\crcr\noalign
{\kern-0.02pt\nointerlineskip}
$\displaystyle{#1}$\crcr}}}
\def\upar#1{\vbox{\ialign{\hfil##\hfil\crcr
$\raise0.3pt\hbox{$\scriptstyle \leftrightarrow$}$\crcr\noalign
{\kern-0.02pt\nointerlineskip}
$\displaystyle{#1}$\crcr}}}
\def\ular#1{\vbox{\ialign{\hfil##\hfil\crcr
$\raise0.3pt\hbox{$\scriptstyle \leftarrow$}$\crcr\noalign
{\kern-0.02pt\nointerlineskip}
$\displaystyle{#1}$\crcr}}}

\def\Tr{\,{\rm Tr }\,}

\def\g5{\gamma_5}

\def\lp1{{\cal L}_{\pi N}^{(1)}}
\def\lp2{{\cal L}_{\pi N}^{(2)}}
\def\lp3{{\cal L}_{\pi N}^{(3)}}

\topskip=0.60truein
\leftskip=0.18truein
\vsize=8.8truein
\hsize=6.5truein
\tolerance 10000
\hfuzz=20pt

\baselineskip 14pt plus 1pt minus 1pt
\pageno=0
\centerline{\bf ASPECTS OF NUCLEON COMPTON SCATTERING}
\vskip 48pt
\centerline{V. Bernard$^1$, N. Kaiser$^2$,
Ulf-G. Mei{\ss}ner$^1$, A. Schmidt$^2$}
\vskip 16pt
\centerline{$^1${\it Centre de Recherches Nucl\'{e}aires et Universit\'{e}
Louis Pasteur de Strasbourg}}
\centerline{\it Physique Th\'{e}orique,
BP 20Cr, 67037 Strasbourg Cedex 2, France}
\vskip  4pt
\centerline{$^2${\it Physik Department T30,
Technische Universit\"at M\"unchen}}
\centerline{\it
James
Franck Stra{\ss}e,
D-85747 Garching, Germany}
\vskip  4pt
\vskip 12pt
\vskip 1.0in
\centerline{\bf ABSTRACT}
\medskip
\noindent We consider the spin-averaged nucleon forward Compton scattering
amplitude in heavy baryon chiral perturbation theory including all terms to
order ${\cal O} (q^4)$. The chiral prediction for the spin-averaged
forward Compton scattering amplitude is in good agreement with the data
for photon energies $\omega \le 110$ MeV. We also evaluate the nucleon electric
and magnetic Compton polarizabilities to this order and discuss the
uncertainties of the various counter terms entering the chiral expansion
of these quantities.
\medskip
\vfill
\noindent CRN--93/xx  \hfill November 1993
\vskip 12pt
\eject
\baselineskip 14pt plus 1pt minus 1pt
\noindent{\bf I. INTRODUCTION AND SUMMARY}
\bigskip
Compton scattering off the nucleon at low energies offers important information
about the structure of these particles in the non-perturbative regime of QCD.
The spin--averaged forward scattering amplitude for real photons in the nucleon
rest frame can be expanded as a power series in the photon energy $\omega$,
$$\eqalign{ T (\omega ) &= f_1 (\omega^2 ) \,
\vec{\epsilon}_f^{\, *} \cdot \vec{\epsilon}_i \cr
f_1 (\omega^2 ) &= a_0 + a_1 \, \omega^2 + a_2 \, \omega^4 + \ldots \cr}
\eqno(1)$$
where $\vec{\epsilon}_{i,f}$ are the polarization vectors of the initial and
final photon, respectively, and due to crossing symmetry only even powers of
$\omega$ occur. The Taylor coefficients $a_i$ encode the
information about the nucleon structure. The first term in eq.(1), $a_0 = -e^2
Z^2 / 4 \pi m$, dominates as the photon energy  approaches zero, it is only
sensitive to the charge  $Z$ and the mass $m$ of the
particle the photon scatters off (the Thomson limit). The term quadratic in
the energy is equal to the sum of the so-called electric ($\bar\alpha$) and
magnetic ($\bar\beta$) Compton polarizabilities, $a_1 = \bar \alpha + \bar
\beta$.  Corrections of higher order in
$\omega$ start out with the term proportional to $a_2$.
\medskip
In ref.[1], we calculated  the electric and magnetic polarizabilities of the
proton and the neutron in heavy baryon chiral perturbation theory (CHPT)
to order
$q^4$ (with $q$ denoting a small external momentum or meson mass). The chiral
prediction for
the electric polarizabilities $\bar{\alpha}_p$ and  $\bar{\alpha}_n$ are in
good agreement with the available experimental data. Furthermore, we found a
non--analytic term of the type $\ln M_\pi$ (with $M_\pi$ the pion mass) which
cancels parts of the large positive $\Delta(1232)$ contribution to the magnetic
polarizability $\bar{\beta}_p$ (in case of the neutron, the coefficient of the
$\ln M_\pi$ term is somewhat too small). This is a novel effect which allows to
understand the relative smallness of the nucleons' magnetic polarizabilities.
The aim of the present paper is twofold. First, we wish to consider the
spin--averaged forward Compton amplitude as a function of the
photon energy to
investigate the importance of the $\omega^4, \, \omega^6, \, \ldots$ terms and
to confront it with the available experimental data. Second, as already
stressed in ref.[1], we have to present a detailed analysis of the
uncertainties entering the theoretical predictions for the polarizabilities
at order $q^4$.
\medskip
Heavy baryon CHPT is a systematic expansion of low--energy  QCD Green functions
in small external momenta and quark masses. It was previously used in the
calculation of the baryon mass spectrum [2], axial currents in flavor SU(3)
[3], nuclear forces [4] and many other observables (for a review, see ref.[5]).
A systematic analysis of QCD Green functions for nucleons in flavor SU(2)
can be found in
[6]. We will heavily borrow from that paper. We will work in the one loop
approximation [6] which still allows to include all terms of chiral order
$q^4$. The two loop
contributions start at ${\cal O}(q^5)$. The nucleon electromagnetic
polarizabilities belong to the rare class of observables which to leading order
in the chiral expansion are given as pure loop effects [7] but at order $q^4$
counter terms enter. Their coefficients have already been determined in [1]
by either direct fits to existing data for other reactions or by making use
of the resonance saturation hypothesis [8,9]. This will be discussed in more
detail below.
\medskip
The pertinent results of this investigation can be summarized as
follows:
\smallskip
\item{(i)}The spin--averaged forward Compton amplitude for the proton is in
agreement with the data up to photon energies $\omega \simeq M_\pi$.
It is dominated by the quadratic contribution, i.e. to order $q^4$
in the chiral expansion the terms of order $\omega^4$ (and higher)
are small. Similar trends are found for the neutron with the exception
of a too strong curvature at the origin.
\smallskip
\item{(ii)}The electromagnetic polarizabilities are in good agreement
with the data (with the exception of $\bar{\beta}_n$), see eqs.(28)
and (29). We have
discussed the theoretical uncertainties to order $q^4$ and found
that the electric and magnetic polarizabilities can be predicted with
an accuracy of $\pm 2 \cdot 10^{-4}$ fm$^3$ and
$\pm 4 \cdot 10^{-4}$ fm$^3$, respectively. An improved theoretical
prediction is only possible if one can pin down the coefficients of
some counter terms more accurately.
\medskip
The paper is organized as follows. In section 2, we briefly discuss the
effective meson-baryon lagrangian underlying our calculation. In section 3 we
introduce the spin-averaged forward Compton tensor which contains the
information about the amplitude $f_1(\omega^2)$ as well as the electromagnetic
polarizabilities $\bar\alpha$ and $\bar \beta$. Section 3 contains the results
for the spin--averaged forward Compton scattering amplitude. The theoretical
predictions for the electromagnetic polarizabilities are then given in section
 4.

\bigskip
\noindent{\bf 2. EFFECTIVE LAGRANGIAN}
\bigskip
The basic tool to investigate the low--energy behaviour of QCD Green functions
is an effective Lagrangian formulated in terms of the asymptotically observed
fields, the Goldstone bosons of spontaneous chiral symmetry breaking (pions)
and the matter fields (the baryons). The
construction principles for building a string of terms with increasing number
of derivatives are chiral symmetry, lorentz invariance and the pertinent
discrete symmetries of the strong interactions
as well as the systematic chiral
power counting scheme (see [5] and references therein). For our
purpose, it suffices to stress that we work in the one loop approximation
including all terms of order $q^4$. As stressed in [3], $n$ loop graphs start
to contribute at order $q^{2n+1}$ if one inserts  vertices from
the lowest order effective Lagrangian ${\cal L}_{\pi N}^{(1)}$.
Notice that we consider the two flavor case and work in the isospin limit
$m_u = m_d$.
The pion fields are collected in the SU(2) matrix
$$U(x) = \exp[i{\vec \tau}\cdot{\vec \pi}(x) / F] = u^2 (x) \eqno(2)$$
with $F$ the pion decay constant in the chiral limit. The
nucleons are considered as very heavy [2,3,4,5,6].
This allows for a projection onto velocity eigenstates and one therefore can
eliminate the troublesome nucleon mass term (of the Dirac lagrangian for
nucleons) thereby generating a string of vertices of increasing power in
$1/m$, with $m$ the nucleon mass (in the chiral limit).\footnote{$^{1)}$}{Later
on, we will identify $m$ with the physical nucleon mass. This is a consistent
procedure to the order we are working.}
The effective Lagrangian to order ${\cal
O}(q^4)$, where $q$ denotes a genuine small momentum or a meson
(quark) mass, reads (we only exhibit those terms which are actually needed in
the calculation of forward Compton amplitudes) [1]
$$\eqalign{{\cal L}_{\pi N} =   & {\cal L}_{\pi N}^{(1)} +
{\cal L}_{\pi N}^{(2)} + {\cal L}_{\pi N}^{(4)} \cr
{\cal L}_{\pi N}^{(1)} = & {\bar H} ( i v\cdot D + g_A S \cdot u ) H \cr
{\cal L}_{\pi N}^{(2)} = & {\bar H} \biggl\lbrace  -{1\over 2m} D \cdot D +
{1 \over 2m} (v \cdot D)^2 + c_1\Tr \chi_+  + \bigl( c_2 - {g_A^2
\over 8 m} \bigr)  v\cdot u\, v\cdot u \cr & + c_3\,  u\cdot u - {i g_A \over
2m} \lbrace S \cdot D , v \cdot u \rbrace - { i e \over 4 m}
[ S^\mu , S^\nu ] \biggl( (1 + \kappa_v) f^+_{\mu \nu} + {\kappa_s - \kappa_v
        \over 2 } \Tr f^+_{\mu \nu} \biggr)
\biggr\rbrace H \cr
{\cal L}_{\pi N}^{(4)} = &  {\pi \over 4} (\delta \bar \beta_p -
\delta \bar \beta_n)
\bar H f_{\mu \nu}^+ f^{\mu \nu}_+ H
                        + {\pi \over 4} \delta \bar  \beta_n
                        \bar H  H \Tr f_{\mu \nu}^+ f^{\mu \nu}_+  \cr
                       &+ {\pi \over 2} (\delta \bar \alpha_n
 + \delta \bar \beta_n
 - \delta \bar \alpha_p
 - \delta \bar \beta_p)
    \bar H f_{\mu \nu}^+ f^{\lambda \nu}_+ H v^\mu v_\lambda  \cr
                       &- {\pi \over 2} (\delta \bar \alpha_n
 + \delta \bar \beta_n)
    \bar H H \Tr(f_{\mu \nu}^+ f^{\lambda \nu}_+ ) v^\mu v_\lambda
 \cr} \eqno(3)$$
with
$$\eqalign{u_\mu &= i u^\dagger \nabla_\mu U u^\dagger \cr
f_{\mu \nu}^+ &= (\partial_\mu A_\nu - \partial_\nu A_\mu)
(u Zu^\dagger + u^\dagger Zu)
\cr} \eqno(4)$$
where $H$ denotes the heavy nucleon field of charge $Z = (1+ \tau_3)/2$
and anomalous
magnetic moment $\kappa = (\kappa_s + \tau_3 \kappa_v)/2$,
$v_\mu$ the four--velocity of $H$, $S_\mu$ the covariant spin--operator
subject to the constraint $v \cdot S = 0$, $\nabla_\mu$ the covariant
derivative acting on the pions, $D_\mu = \partial_\mu + \Gamma_\mu $ the chiral
covariant derivative for nucleons and we adhere to the notations of ref.[6].
The superscripts (1,2,4) denote the chiral power.  The lowest order effective
Lagrangian is of order ${\cal O}(q)$. The one loop contribution is suppressed
with respect to the tree level by $q^2$ thus contributing at ${\cal O}(q^3)$.
In addition, there are one loop diagrams with exactly one insertion from
${\cal L}_{\pi N}^{(2)}$. These are of order $q^4$.
Finally, there are contact terms of order $q^2$ and $q^4$ with coefficients
not fixed by chiral symmetry. For the case at hand ${\cal L}_{\pi N}^{(3)}$
does not have to be considered explicitely. To order $q^4$ its vertices enter
only tree diagrams and it therefore contributes to
the polynomial part of the
amplitude. The most general polynomial piece at order $q^4$ is, however,
already given by the contact vertices of ${\cal L}_{\pi N}^{(4)}$. Notice that
some coefficients in ${\cal L}_{\pi N}^{(2)}$ related to the $\gamma \gamma N
N$ and $\gamma \pi N N$ vertices are fixed from the relativistic theory
by the low-energy theorems for Compton scattering ($a_0 = -e^2 Z^2 / 4\pi m$)
and neutral pion photoproduction, respectively. This is discussed in some
detail in ref.[6]. The unknown coefficients we have to determine are $c_1$,
$c_2$ and $c_3$ characterizing a higher derivative $\pi\pi NN$ vertex as well
as the four low-energy constants $\delta
\bar{\alpha}_p$, $\delta \bar{\alpha}_n$, $\delta \bar{\beta}_p$ and $\delta
\bar{\beta}_n$ from ${\cal L}_{\pi N}^{(4)}$.
We have not exhibited the standard meson Lagrangian ${\cal L}_{\pi \pi}^{(2)}$,

To calculate all terms up-to-and-including order $q^4$, we have to evaluate all
one loop graphs with insertions from ${\cal L}_{\pi N}^{(1)}$ and those
with exactly one
insertion from ${\cal L}_{\pi N}^{(2)}$. While the former scale as  $q^3$, the
latter constitute the new contributions of order $q^4$. Furthermore, there are
the tree diagrams related to ${\cal L}_{\pi N}^{(4)}$ which are also new. It is
worth to stress that in the one loop diagrams involving ${\cal L}_{\pi
N}^{(2)}$ the anomalous magnetic moment of the nucleon appears, since the
photon-nucleon vertex is expanded in the external momentum.  We will come
back to this point later on.  Finally, it is
mandatory to expand the leading order ${\cal O}(q)$ effective vertices
as well as the nucleon propagator  to include all relativistic
corrections of order
$1 / m$. The resulting Feynman rules for the pertinent vertices and propagators
are shown in fig.1. Note that we have left out all vertices which give zero
contribution either due to spin-averaging or due to isospin algebra.

In eq.(3) we have expressed the effective lagrangian in terms of the physical
parameters $m, \kappa_s, \kappa_v$ and so on. For the loop diagrams this is
legitimate to the order we are working. Concerning tree diagrams which are
responsible for the Thomson amplitude $a_0 = -e^2 Z^2 / 4 \pi m$, the term
proportional to $c_1$ will shift the nucleon mass in the chiral limit
by $-4 c_1 M_\pi^2$ to  the physical nucleon mass at this order.
\vfill
\eject
\noindent{\bf 3. SPIN-AVERAGED FORWARD COMPTON TENSOR}  \bigskip
The object one has to study in order to determine the Compton amplitude
$f_1(\omega^2)$ and the electromagnetic polarizabilities is the spin-averaged
Compton tensor in forward direction $\Theta_{\mu\nu}$,
$$\eqalign{\Theta_{\mu\nu} & = {e^2\over 4} \Tr[ ( 1 + \barre v) T_{\mu\nu}
(v,k)] \cr & = A(\omega) \, g_{\mu\nu}  + B(\omega) \, k_\mu k_\nu + C(\omega)
\, ( k_\mu v_\nu + v_\mu k_\nu) + D(\omega) \, v_\mu v_\nu \cr}\eqno(5)$$
where $v$ and $k$ denote the nucleon four-velocity  ($v^2 = 1$) and the photon
momentum ($k^2= 0$) and $\omega = v\cdot k$. $T_{\mu\nu}(v,k)$ is the
Fourier-transformed nucleon matrix element of two time-ordered
electromagnetic currents,
$$T_{\mu\nu}(v,k) = \int d^4x e^{ik\cdot x} <N(v)|{\cal T} J^{em}_\mu(x)\,
J^{em}_\nu(0)|N(v)> \eqno(6)$$
Gauge invariance $\Theta_{\mu\nu} \, k^\nu = 0$ implies $C(\omega) = -
A(\omega)/\omega$ and $D(\omega) = 0$ and therefore all information is
contained in the two crosssing even functions $A(\omega)= A(-\omega)$ and
$B(\omega) = B(-\omega)$. This fact
allows us to choose the "Coulomb gauge" $\epsilon \cdot v = 0$ for the
polarization vector $\epsilon$ of the photon when calculating the auxiliary
quantity $\epsilon^\mu\,\Theta_{\mu\nu}\,\epsilon^\nu = A(\omega) \, \epsilon^2
+ B(\omega)\, (\epsilon\cdot k)^2$. The gauge $\epsilon \cdot v = 0$ is very
convenient since it
leads to a drastic simplification in our loop calculation. In
this gauge the leading order photon nucleon vertex vanishes and therefore many
diagrams. The function $A(\omega)$ is directly proportional to the
spin-averaged forward Compton amplitude $f_1(\omega^2) $
introduced in eq.(1), namely,
$$A(\omega) = - 4 \pi \, f_1(\omega^2) \eqno(7)$$
The second function $B(\omega)$  is not a physical amplitude, it only
serves to calculate the magnetic polarizability.

The electric and magnetic Compton polarizabilities are defined as follows:
$$\bar \alpha + \bar \beta = - {A''(0)\over 8 \pi} \,, \qquad \quad \bar
\beta = - {B(0) \over 4 \pi}  \eqno(8)$$
where the prime denotes differentiation with respect to $\omega$.

Our task is to calulate all contributions to $A(\omega)$ and $B(\omega)$ coming
from CHPT at order $q^4$. These are all one loop contributions with vertices
from ${\cal L}_{\pi N}^{(1)}$ and those with exactly one vertex fron ${\cal
L}^{(2)}_{\pi N} $ as well as the most general polynomial contribution at this
order. The latter has the form $A(\omega)^{pol} = e^2 Z^2/m - 4 \pi (\delta
\bar \alpha
+ \delta \bar \beta)\, \omega^2$ and $B(\omega)^{pol} = - 4 \pi \delta\bar
\beta$ where the     constant term in $A(\omega)^{pol}$ is fixed by the low
energy theorem for Compton scattering. Later we will determine the polynomial
coefficients $\delta \bar \alpha,\, \delta \bar \beta$ by identifying
them with
the resonance contributions to the nucleon electromagnetic polarizabilities. In
making this identification we have excluded any contributions from the nucleon
pole graphs to the polarizabilities. This prescription is in accordance with
the analysis of
the Compton scattering experiments where terms coming from the anomalous
magnetic moment (nucleon pole graph) are explicitely separated from the ones of
the polarizabilities by the use of the Powell cross section formula [15].
\bigskip
\noindent{\bf 4. FORWARD COMPTON AMPLITUDE}
\bigskip
We wish to discuss the spin--averaged forward Compton amplitude $A(\omega) =
-4\pi f_1(\omega)$ for real  photons ($k^2=0$) scattering off protons or
neutrons.  Making use of a once--subtracted dispersion
relation and the optical theorem allows to express $A(\omega)$ in terms of the
total photo--nucleon absorption cross section $\sigma_{\rm tot}(\omega)$ as
$$\eqalign{
{\rm Re}\,A(\omega) &= {e^2 Z^2 \over m} - {2 \omega^2 \over \pi}\, {\cal P}
\int_{\omega_0}^\infty \, d\omega' {\sigma_{\rm tot}(\omega') \over {\omega'}^2
- \omega^2}  \cr
{\rm Im}\,A(\omega) &= - \omega \, \sigma_{\rm tot}(\omega) \cr} \eqno(9)$$
for the proton and the neutron. The threshold energy for single pion
photoproduction is given by
$$\omega_0 = M_\pi(1 + M_\pi /2 m) \,\, . \eqno(10)$$

Working in
the Coulomb gauge $\epsilon \cdot v =0$, one has to calculate 9 irreducible
one loop graphs with insertions from ${\cal L}_{\pi N}^{(1)}$ and 67 one loop
diagrams with one insertion from ${\cal L}_{\pi N}^{(2)}$. Furthermore there
are the contact term graphs stemming from ${\cal L}_{\pi N}^{(4)}$, which give
rise to the most general polynomial contribution at order $q^4$ to $A(\omega)$
and $B(\omega)$, namely, $A(\omega)^{pol} = e^2 Z^2/m - 4\pi (\delta \bar
\alpha + \delta \bar \beta) \omega^2, \quad B(\omega)^{pol} = - 4\pi \delta
\bar \beta$.

Due to the simple structure of the loop integrals in the heavy mass
formulation, one can give $A_{p,n}(\omega)$ in closed analytic form. For the
proton it reads
$$\eqalign{ A_p (\omega) = & {e^2 \over m} - 4 \pi  (\bar{\alpha}+\bar
{\beta})_p \omega^2 \cr
& + {e^2 g_A^2 M_\pi \over 8 \pi F_\pi^2} \biggl\lbrace-{3
\over 2} -{1 \over z^2} + \bigl(1+{1 \over z^2}\bigr) \sqrt{1 - z^2} +
{1 \over z} \arcsin z + {11 \over 24} z^2 \biggr\rbrace \cr
 & + {e^2 g_A^2 M_\pi^2 \over 8 \pi^2 m F_\pi^2} \biggl\lbrace -{5\over 2}
+{z^2\over 2}(3 \kappa_s + 11) + \biggl[{2 \over z} - (6+\kappa_s) z +
{z \over 1 - z^2}\biggr]  \sqrt{1 - z^2} \arcsin z \cr &  \qquad \qquad \quad
+ {1 \over 2} \bigl({1 \over z^2} - \kappa_s \bigr) \arcsin^2 z \biggr\rbrace
\cr} \eqno(11)$$
with $z = \omega / M_\pi$. Similarly, one has for the neutron
$$\eqalign{ A_n (\omega) = & - 4 \pi (\bar{\alpha}+\bar{\beta})_n \omega^2
+ {e^2 g_A^2 M_\pi \over 8 \pi F_\pi^2} \biggl\lbrace -{3 \over 2} -{1 \over
z^2} + \bigl(1+{1 \over z^2}\bigr) \sqrt{1 - z^2} + {1 \over z} \arcsin z +
{11 \over 24} z^2 \biggr\rbrace \cr
 & + {e^2 g_A^2 M_\pi^2 \over 8 \pi^2 m F_\pi^2}
\biggl\lbrace {1\over 2} +{z^2\over 6}(7- 9 \kappa_s) + \biggl[
(\kappa_s - 2) z + {z \over 1 - z^2}\biggr] \sqrt{1 - z^2} \arcsin z
\cr & \qquad\qquad \quad  + {1 \over 2} \bigl( \kappa_s - {1 \over z^2} \bigr)
\arcsin^2 z \biggr\rbrace \cr} \eqno(12)$$
Notice that the expressions for $A_{p,n}(\omega)$ diverge at $\omega = M_\pi$.
This is an artefact of the heavy mass expansion. The realistic branch point
coincides with the opening of the one--pion channel as given above. To cure
this, let us introduce the variable
$$\zeta = {z \over 1 + M_\pi/2m} = {\omega \over M_\pi ( 1 + M_\pi /2m )}  =
{\omega \over \omega_0} \, . \eqno(13)$$
If one now rewrites $A_{p,n}(\omega)$ in terms of $\zeta$, the branch point
sits at its proper location and $A_{p,n} (\zeta = 1)$ is finite. We have
$$\eqalign{A_{p,n} (\omega) & =  {e^2 \over 2m} (1 \pm 1)
- 4 \pi (\bar{\alpha}+\bar{\beta})_{p,n} \omega^2 \cr
& + {e^2 g_A^2 M_\pi \over 8 \pi F_\pi^2} \biggl\lbrace -{3 \over 2} -{1 \over
 \zeta^2} + \bigl(1+{1 \over \zeta^2}\bigr) \sqrt{1 - \zeta^2} +
{1 \over \zeta} \arcsin\zeta + {11 \over 24} \zeta^2 \biggr\rbrace \cr
 & + {e^2 g_A^2 M_\pi^2 \over 8 \pi^2 m F_\pi^2} \biggl\lbrace -1 +{10\over 3}
\zeta^2 + \biggl[{1 \over \zeta} - 4 \zeta +
{\zeta \over 1 - \zeta^2}\biggr] \sqrt{1 - \zeta^2} \arcsin\zeta \cr
 & + \pi \biggl[ {1 \over \zeta^2}-{1 \over 2 \zeta} \arcsin \zeta
+ {11 \over 24}\zeta^2 - {(1-\zeta^2)^2 + 1 \over 2 \zeta^2 \sqrt{1-\zeta^2}}
\biggr]  \pm \biggl[ -{3 \over 2} + \zeta^2\bigl( {3\over 2} \kappa_s +
{13\over 6} \bigr) \cr & + \bigl({1 \over \zeta} -(2+\kappa_s)
\zeta\bigr) \sqrt{1 -\zeta^2} \arcsin\zeta + {1 \over 2} \bigl({1 \over
\zeta^2}-\kappa_s \bigr) \arcsin^2\zeta \biggr] \biggr\rbrace \cr} \eqno(14)$$
where the '+/-' sign refers to the proton/neutron, respectively. The proper
analytic continuation above the branch point $\zeta = 1$ is obtained through
the substitutions $\sqrt{1-\zeta^2} = -i \sqrt{\zeta^2 -1} $ and $\arcsin \zeta
 = \pi/2 + i \ln(\zeta + \sqrt{\zeta^2-1}) $.
The expressions given in eq.(14) differ from the ones in eqs.(11,12) only by
terms of
order $q^5$ (and higher) and are thus equivalent to the order we are working.
We should stress that in the relativistic formulation of baryon CHPT such
problems concerning the branch point do not occur [10,11].
In the heavy mass
formulation we encountered this problem since the branch point $\omega_0$
itself has an expansion in $1/m$ and is thus different in CHPT at order $q^3$
and $q^4$.

We now present our numerical results. We always use the
Goldberger-Treiman relation to express $g_A / F_\pi$ as $g_{\pi N} / m$ ,
with $g_{\pi N} = 13.40$ the strong pion--nucleon coupling constant.
We also use $\kappa_s = -0.12$, $m = 938.27$ MeV and
$e^2/4 \pi = 1/137.036$.
In fig.2a, we show the real part of $A_p (\omega)$ normalized to the
Thomson limit, $e^2/m  = 3.02$   $\mu {\rm b \, GeV}^{-1}$,
for the $z$ and
$\zeta$ expansions in comparison to the data [12] for the central value
of $(\bar{\alpha}+\bar
\beta )_p = 14.0 \cdot
10^{-4}$ fm$^3$ of ref.[1].
Up to $\omega \simeq 100$ MeV, the agreement of the prediction with
the data is good. The corrections of order $\omega^4$ (and higher)
are fairly small as shown in fig.2b. One also recognizes the unphysical
behaviour of the $z$--expansion around $\omega = M_\pi$. Similar
statements hold for the neutron exhibited in fig.3 using the value
$(\bar{\alpha}+\bar\beta )_n = 21.2 \cdot
10^{-4}$ fm$^3$ of ref.[1].
Using the deuteron data of Armstrong et al.[13], one finds at $\omega
= 100$ MeV, $A_n / A_p = -0.35$ compared to the theoretical prediction
of $-0.47$. The difference is mostly due to the too large sum of the
electric and magnetic polarizabilities of the neutron.

For the proton, we have also calculated the real and imaginary parts
of $A(\omega)$ for $\zeta > 1$. The imaginary part starts out negative
as it should but becomes positive at $\omega \simeq 180$ MeV. This is
due to the truncation of the chiral expansion and can only be overcome
by a more accurate higher order calculation. Consequently, the real
part (normalized to the Thomson limit) stays rather flat after the
branch point as shown in fig.4.
\bigskip
\noindent{\bf 5. ELECTROMAGNETIC POLARIZABILITIES}
\bigskip
In [1], we derived the following formulae for the electric and magnetic
polarizabilities of the proton and the neutron ($i = p,n$)
$$\eqalign{
  \bar{\alpha}_i&= {5 C g_A^2 \over 4 M_\pi} + {C \over \pi} \biggl[ \bigl(
{x_i
g_A^2 \over m} - c_2 \bigr) \ln {M_\pi \over \lambda} + {1 \over 4} \bigl(
{y_i g_A^2
\over 2 m} - 6 c_2 + c^+ \bigr) \biggr] +\delta
\bar{\alpha}_i^r(\lambda) \, ,   \cr
  \bar{\beta}_i&= { C g_A^2 \over  8 M_\pi} + {C \over \pi}  \biggl[ \bigl( {3
x_i' g_A^2 \over m} - c_2 \bigr) \ln {M_\pi \over \lambda} + {1 \over 4}
\bigl( {y_i' g_A^2
\over  m} + 2 c_2 - c^+ \bigr) \biggr] + \delta \bar{\beta}_i^r (\lambda)
\, .  \cr} \eqno(15)$$
with
$$\eqalign{
 C  &= {e^2   \over 96 \pi^2 F_\pi^2}\, \, = 4.36 \cdot 10^{-4} \, {\rm fm}^2
\, , \cr
  x_p&= 9 \, , \quad x_n = 3 \, , \quad y_p = 71 \, , \quad y_n = 39 \, ,
\cr
  x_p'&= 3 + \kappa_s \, , \quad x_n' = 1 - \kappa_s \, , \quad y_p' = {37
\over
2} + 6 \kappa_s \, , \quad y_n' = {13 \over 2} - 6 \kappa_s \, , \cr
  c^+&= -8 c_1 + 4 c_2 + 4 c_3 - {g_A^2 \over 2 m} \, \, \, . \cr}
\eqno(16)$$
Here,  $\lambda$ is the scale introduced in dimensional regularization. The
physical $\bar \alpha_i$ and $\bar \beta_i$
are of course independent of this scale since
the renormalized counter terms $\delta \bar \alpha_i^r(\lambda)$
and $\delta \bar \beta_i^r(\lambda)$
cancel  the logarithmic $\lambda$--dependence of the loop contribution.
The corresponding renormalization prescription reads:
$$
\delta \bar \alpha_i =  {e^2 L \over 6\pi F_\pi^2} \biggl( c_2 -{x_i g_A^2
\over m} \biggr) + \delta \bar \alpha^r_i(\lambda), \qquad
\delta \bar \beta_i =  {e^2 L \over 6\pi F_\pi^2} \biggl( c_2 -{3x'_i g_A^2
\over m} \biggr) + \delta \bar \beta^r_i(\lambda) \eqno(17)$$
with
$$L = {\lambda^{d-4} \over 16 \pi^2 }\biggl[ {1\over d-4} + {1 \over 2}
(\gamma_E - 1 - \ln 4\pi) \biggr] \eqno(18)$$
The first term on
the r.h.s. of eq.(15) is, of course, the result at order $q^3$ [6,7,10].
The results shown in eqs.(15) have the following structure.
Besides the leading $1/M_\pi$ term [6,7,10],
${\cal O}(q^4)$ contributions from the loops have a $\ln M_\pi$ and a constant
piece $\sim M_\pi^0$. As a check one can recover the coefficient of the
$\ln M_\pi$ term form the relativistic calculation [1,2] if one sets
the new low energy constants $c_i$ and $\kappa_{s,v} = 0$. In that
case only the $1/m$ corrections of the relativistic Dirac formulation
are treated and one necessarily reproduces the corresponding
non--analytic (logarithmic) term of this approach.
The term proportional to $c_2 \, \ln M_\pi$ in eqs.(15) represents the  effect
of (pion) loops with intermediate $\Delta (1232)$ states [14] consistently
truncated at order $q^4$. We should stress that the decomposition of
the loop and counter term pieces at order $q^4$ has, of course, no deeper
physical meaning but will serve us to separate the uncertainties
stemming from the coefficients accompanying the various contact terms.
Notice that from now on we will omit the superscript '$r$' on
$\delta \bar \alpha_i^r(\lambda)$ and $\delta \bar \beta_i^r(\lambda)$
appearing in eqs.(15).

The numerical results for the various polarizabilities now depend on the
knowledge of the contact terms $c_2$, $c^+$,$\delta \bar \alpha_i(\lambda)$ and
$\delta \bar \beta_i(\lambda)$ for a given choice of the scale $\lambda$.
As already stated, ideally the $\lambda$--dependence from the loops is
cancelled by the one from the corresponding contact terms, as detailed in
[16], when one is able to fit all low--energy constants from phenomenology.
We are not in that fortunate position but in some cases must resort to the
resonance saturation hypothesis [8,9] to estimate some of the constants. In
that case,  the actual value of a certain low--energy constant is given as a
sum of a dominating resonance contribution at the scale $\lambda$ equal to some
resonance scale. This problem
will eventually be cured when sufficiently many accurate low--energy data in
the baryon sector will be available. Clearly, the contact terms we are dealing
with fall into two categories.  While $c_2$ and $c^+$ only enter via higher
order (in $1/m$) vertices in the loop diagrams,
$\delta \bar \alpha_i(\lambda)$ and $\delta \bar \beta_i(\lambda)$ are
"genuine" new contact terms of order $q^4$. We will
therefore discuss the sensitivity on these separately. Note, however, that only
the total result of order $q^4$ is of physical relevance.

The constant $c^+$ can be related to the isospin--even S--wave pion--nucleon
scattering length $a^+$ [17],
$$c^+ = {256 \pi^2 F_\pi^4 a^+ - 3 g_A^2 M_\pi^3 \over 32 \pi M_\pi^2 F_\pi^2
( 1 - M_\pi /m )} = - 0.28 \ldots - 0.42 \, {\rm fm}
\eqno(19)$$
for  $a^+ = -0.83 \pm
0.38 \cdot 10^{-2} / M_\pi$ with
    $M_\pi = 139.57$ MeV the (charged) pion mass. Inserting the value of
$c^+$ as given from eqs.(19) into eqs.(15,16), one finds that the contributions
proportional to $c^+$ to the electromagnetic polarizabilities are negligible
since they are less then $0.15 \cdot 10^{-4}$ fm$^3$
in magnitude. If we take only the leading order relation $c^+ = F_\pi^2 a^+ /
8\pi M_\pi^2 $,
which is allowed to the order we are working, then the $c^+$
contribution to the electromagnetic polarizabilities is even smaller and only
$0.06 \cdot 10^{-4}$ fm$^3$ in magnitude. To determine $c_2$, we make use
of the resonance saturation hypothesis and get contributions from the
$\Delta(1232)$ as well as the Roper $N^* (1440)$,
$$\eqalign{
c_2^\Delta &= {g_A^2 m \over 4 m_\Delta^2} \biggl[ {m_\Delta + m \over m_\Delta
- m} - 4 Z^2 \biggr] \cr
c_2^{N^*} &= {R g_A^2 m \over 8 (m_{N^*}^2 - m^2)} \cr} \eqno(20)$$
where $Z$ parametrizes the off--shell behaviour of the spin--3/2
Rarita-Schwinger field in the $\Delta N \pi$--vertex and
we have used $g_{\pi N \Delta} = 3g_{\pi N} / \sqrt{2}$ as well as
$g_{\pi N N^*} = \sqrt{R} \,  g_{\pi N} / 2 $.
Empirically, $Z$ is not
known very accurately, $-0.8 \le Z \le 0.3$ [18], and $R$ varies from 0.25 to
1.\footnote{$^{2)}$}{In ref.[19], a narrower range for $Z$ is given under the
assumption of $g_2 = 0$ which seems to be excluded by the data as stressed
in [18]. We therefore use the wider range of $Z$ given in [18].}
Notice that the Roper contribution is more than one order of magnitude smaller
than the one from the $\Delta$. If one treats the $\Delta$ non-relativistically
(isobar model), one arrives at
$$c_2 = {g_A^2 \over 2 ( m_\Delta -m )} = 0.59 \, {\rm fm}   \eqno(21)$$
for   $g_A = 1.328$ by making use of the Goldberger--Treiman relation.
Altogether, $c_2$ varies between 0.4 and 0.6 fm. The resulting
polarizabilities depend only weakly on the actual value of $c_2$ as shown in
fig.5. (for $\lambda = 1.232$ GeV). As discussed before, there is some
spurious sensitivity on the value of $\lambda$ as shown in fig.6.
The corresponding
bands refer to the choice of $g_A =1.26$ or $g_A = 1.328$ which are equivalent
to the order we are working. While the neutron polarizabilities are quite
insensitive to the choice of $\lambda$, the much larger coefficients $x_p$ and
$x_p'$ induce some scale dependence for the proton case. Since most of the
counterterms are in fact given by $\Delta$ exchange, we have chosen in [1]
$\lambda = m_\Delta$ (which gives our best values).

The $\Delta(1232)$ enters prominently in the determination of the four
low-energy constants  from ${\cal L}_{\pi N}^{(4)}$.
Therefore, we will determine these coefficients at the scale
$\lambda = m_\Delta$.
In particular, one gets
a sizeable contribution to the magnetic polarizabilties due to the strong
$N\Delta$ M1 transition.
A crude estimate of this has been given in ref.[20]
by integrating the M1 part of the total photoproduction cross section for
single pion photoproduction over the resonance region,
$$\delta \bar{\beta}_p^\Delta (m_\Delta) = {1 \over 2 \pi^2} \int {d \omega
\over \omega^2} \, \sigma^{M1}(\omega) = 7.0 \cdot 10^{-4} \, {\rm fm}^3
  \eqno(22)$$
However, this number is afflicted with a large uncertainty.
If one simply uses the Born
diagrams with an intermediate point--like
$\Delta$, one finds
$$\delta \bar{\beta}_p^\Delta (m_\Delta) = {e^2 g_1^2 \over 18\pi m^2
m_\Delta^2 } \biggl\lbrace {m^2_\Delta + m_\Delta m + m^2  \over m_\Delta - m}
- 4Y \bigl[  m_\Delta ( 2Y+1) + m(Y+1)\bigr] \biggr\rbrace  \eqno(23) $$
with $g_1$ the strength of the $\gamma N \Delta$ coupling and the off--shell
parameter $Y$ is related to the electromagnetic interaction ${\cal L}^1_{\gamma
N \Delta}$ (see ref.[18] for more details on this). These parameters are not
very well determined, a best fit to
multipole data for pion photoproduction leads
to $3.94 \le g_1 \le 5.30$ and $-0.75 \le Y \le 1.67$ [18]. The recent PDG
tables give $3.5  \le g_1 \le 7.5$ [21]. For comparison, SU(4) and chiral
soliton models give $g_1 = \kappa^* = 5.0$ which was used e.g. in [10] together
with $Y = -1/4$ to estimate the $\Delta$ contribution at order $q^4$.
Inspection of eq.(23) reveals that the large positive values of $Y$ lead to
very large negative contributions from the $\Delta$ in plain contradiction to
the dispersive estimate of eq.(22). If one, however, assumes an universal
off--shell parameter for the strong and electromagnetic $N \Delta$ transitions,
it is plausible to set $-0.8 \le Y \le 0.3$. In that case,
$\delta \bar{\beta}^\Delta$ is
always positive and varies between 14. and 0.5 as shown in fig.7. A
conservative estimate therefore is
$$\delta \bar{\beta}_p^\Delta (m_\Delta) =
\delta \bar{\beta}_n^\Delta (m_\Delta) \simeq (7.0 \pm 7.0)
\cdot 10^{-4}
\, {\rm fm}^3  \eqno(24)$$
invoking isospin symmetry. Clearly, the large range in the
value for $\delta \bar{\beta}^\Delta$
is unsatisfactory and induces a major uncertainty in
the determination of the corresponding counterterms. In [1], we choose the
central value of eq.(24) as our best determination. In the framework of a
non--relativistic calculation (isobar model), the corresponding Born term
which generates  $\delta \bar{\beta}^\Delta$
does not depend on any off--shell parameter
and one finds $e^2 g_1^2 /( 18 \pi m^2 (m_\Delta - m)) = 12 \cdot 10^{-4}$
fm$^3$ (for $g_1 = 5$).
In ref.[22], the $\Delta(1232)$ was included in the effective field
theory as a dynamical degree of freedom and treated non--relativistically (like
the nucleon).\footnote{$^{3)}$}{Notice that such an approach does not have a
consistent chiral power counting as shown in [23]. It might be justified in the
limit of an infinite number of colors where the nucleon and the $\Delta$ are
degenerate in mass.}
It has been argued in [22] that
the $\Delta$ Born graphs have to be calculated at
the off--shell point $\omega = 0$. This effect can
reduce the large $\delta \bar{\beta}^\Delta$
 by almost an order of magnitude. This is reminiscent of the off--shell
dependence discussed before. It was already pointed out in ref.[10]
that a relativistic treatment of the $\Delta(1232)$ also induces a finite
electric polarizability at order $q^4$. This contribution depends strongly
on the $\gamma \Delta N$ couplings $g_1$ and $g_2$ as well as the two
off--shell parameters $ X, Y$,
$$\eqalign{ \delta
\bar\alpha^\Delta(m_\Delta) = & {e^2  \over 18 \pi m^2 m_\Delta^2}\biggl\lbrace
g_1^2 \biggl[   - {m_\Delta^2 \over m_\Delta + m} + 4Y \bigl(m_\Delta(1 + 2Y)
- m Y)\bigr) \biggr] \cr & + g_1 g_2 \biggl[ {m_\Delta ( m - m_\Delta ) \over
2(m_\Delta +m
) } + (X+Y + 4XY) m_\Delta - Y(1+2X) m \biggr] \cr & + {g_2^2 \over 4} \biggl[
 - { 4m_\Delta^2 + m_\Delta m + m^2 \over 4(m_\Delta +m) } + X(1+2X) m_\Delta
- X(1+X) m \biggr] \bigg\rbrace \cr} \eqno(25)$$
The resulting numbers for $\delta \bar{\alpha}^\Delta$ vary
between $-6 \cdot 10^{-4}$ fm$^3$ and $+4 \cdot 10^{-4}$ fm$^3$
for the ranges $-0.8 \le X,Y \le
0.4$, $4 \le g_1 \le 5$ and $4.5 \le g_2 \le 9.5$.
In
the absence of more stringent bounds on these parameters, we            will
set $\delta\bar{\alpha}_{p,n}^\Delta(m_\Delta) = 0$
and assign the theoretical predictions for the
electric polarizabilies an error of $\pm 2 \cdot 10^{-4}$ fm$^3$ accordingly.

Another contribution to the coefficients $\delta \bar{\alpha}_i(\lambda)$ and
$\delta \bar{\beta}_i(\lambda)$ comes from loops involving charged kaons [24].
Since
we are working in SU(2), the kaons and etas are frozen out and effectively
give some finite contact terms. To improve the estimate given in [1],
we include the average nucleon--$\Lambda , \Sigma^0$ mass splitting,
$$ \Delta = {1 \over 2} (m_\Lambda + m_{\Sigma^0} - 2m) = 216 \, {\rm MeV}
\eqno(26)$$
For the electric and magnetic polarizabilities, this leads to
$$\eqalign{
\delta \bar{\alpha}_p^K (m_\Delta) &= { C \over 6 \pi}
(D^2+3F^2) \, F_1(M_K, \Delta) , \quad
\delta \bar{\alpha}_n^K (m_\Delta) = {C \over 4 \pi}
(D-F)^2 \, F_1 (M_K , \Delta )  \cr
\delta \bar{\beta}_p^K (m_\Delta) &= { C \over 6 \pi}
(D^2+3F^2) \, F_2(M_K, \Delta) , \quad
\delta \bar{\beta}_n^K (m_\Delta) = {C \over 4 \pi}
(D-F)^2 \, F_2 (M_K , \Delta )  \cr
F_1 (M_K, \Delta ) &=
{9 \Delta \over \Delta^2 - M_K^2} +
{10 M_K^2 - \Delta^2 \over (M_K^2 - \Delta^2 )^{3/2} }\arccos{\Delta \over M_K}
\cr F_2 (M_K, \Delta ) &= {1 \over \sqrt{M_K^2 - \Delta^2}} \arccos {\Delta
\over M_K} \cr}
\eqno(27)$$
If one sets $m_\Lambda = m_{\Sigma^0} = m$ (which is consistent to the order
we are working),
one recovers eq.(14) of ref.[1]. With $D = 0.8$, $F = 0.5$ and
$M_K = 495$ MeV one finds
$\delta \bar{\alpha}_p^K (m_\Delta) = 1.31 \cdot 10^{-4}\, {\rm fm}^3$ and
$\delta \bar{\alpha}_n^K (m_\Delta) = 0.13 \cdot 10^{-4}\, {\rm fm}^3$.
The corresponding numbers for the kaon contributions to the magnetic
polarizabilities are a factor $F_2 / F_1 = 0.12$ smaller. In the case of
mass degeneracy for baryons this factor is exactly $1/10$ [24].
The values based on (27) might, however, considerably overestimate the
kaon loop contribution. Integrating
e.g. the data from ref.[27] for $\gamma
p \to K \Lambda,K \Sigma^0$, one gets a much smaller contribution since
the typical cross sections are of the order of a few $\mu$barn. This
points towards the importance of a better understanding of SU(3)
breaking effects. At present, we can not offer a solution to resolve
this discrepancy.\footnote{$^{4)}$}{We are
grateful to Anatoly L'vov for drawing our
attention to this problem.}

We now are in the position to give a prediction for the electromagnetic
polarizabilities. As our central (best) values we take the ones from
[1]. The main sources of uncertainty stem from the choice of the value
for $g_A$, the scale $\lambda$ and the kaon and $\Delta$ contributions
to the various low--energy constants. Adding these in quadrature leads
to
$$\eqalign{
\bar{\alpha}_p &= (10.5 \pm 2.0)
\cdot 10^{-4} {\rm fm}^3 \, \quad
\bar{\beta}_p = (3.5 \pm 3.6) \cdot 10^{-4} {\rm fm}^3 \, \cr
\bar{\alpha}_n &= (13.4 \pm 1.5) \cdot 10^{-4} {\rm fm}^3 \, \quad
\bar{\beta}_p = (7.8 \pm 3.6) \cdot 10^{-4} {\rm fm}^3 \, \cr}
\eqno(28)$$
which with the exception of $\bar{\beta}_n$
agree with the empirical data [25]
$$\eqalign{
\bar \alpha_p
&= (10.4 \pm 0.6)
\cdot 10^{-4} {\rm fm}^3
\, , \quad
\bar \beta_p   = (3.8 \mp 0.6)
\cdot 10^{-4} {\rm fm}^3
\cr
  \bar \alpha_n  &= (12.3  \pm 1.3)
\cdot 10^{-4} {\rm fm}^3 \, , \quad
\bar \beta_n   = (3.5 \mp 1.3)
\cdot 10^{-4} {\rm fm}^3 \, .  \cr}
\eqno(29)$$
making use of                the dispersion sum rules [12,26]
$(\bar \alpha+ \bar \beta)_p = (14.2 \pm 0.3) \cdot 10^{-4}$ fm$^3$ and
$(\bar \alpha+ \bar \beta)_n = (15.8 \pm 0.5) \cdot 10^{-4}$
fm$^3$.\footnote{$^{5)}$}{Notice
that the uncertainty on the sum rule value for the neutron
is presumably underestimated since one has to use
deuteron data to extract the photon-neutron cross section.}
Notice that we have added the systematic and statistical errors of
the empirical determinations in quadrature.
Clearly, an independent determination of the electric and magnetic
nucleon polarizabilities would be needed to further tighten the
empirical bounds on these fundamental quantities. This was also
stressed in ref.[22]. It is worth to point out that the uncertainties
given in (28) do not include effects of two (and higher) loops which
start out at order $q^5$. We do not expect these to alter the
prediction for the electric polarizabilities significantly [1].
Such an investigation is underway but goes
beyond the scope of this paper.
Notice also that at present the theoretical
uncertainties are larger than the experimental ones (if one imposes
the sum rules for $(\bar \alpha + \bar \beta)$).
That there is more spread in
the empirical numbers when the dispersion sum rules are not used
can e.g. be seen
in the paper of Federspiel et al. in ref.[25].
\bigskip
\bigskip
{\bf REFERENCES} \bigskip
\item{1.}V. Bernard, N. Kaiser, A. Schmidt
and Ulf-G. Mei{\ss}ner, {\it Phys. Lett.\/} {\bf B}, in print.
\smallskip
\item{2.}J. Gasser and H. Leutwyler, {\it Phys. Rep.\/}
{\bf 87} (1982) 77.
\smallskip
\item{3.}E. Jenkins and A.V. Manohar, {\it Phys. Lett.\/} {\bf B255} (1991)
558.
\smallskip
\item{4.}S. Weinberg, {\it Nucl. Phys.\/} {\bf B363} (1991) 3.
\smallskip
\item{5.}Ulf-G. Mei{\ss}ner,
{\it Rep. Prog. Phys.\/} {\bf 56} (1993) 903.
\smallskip
\item{6.}V. Bernard, N. Kaiser, J. Kambor
and Ulf-G. Mei{\ss}ner, {\it Nucl. Phys.\/} {\bf B388} (1992) 315.
\smallskip
\item{7.}V. Bernard, N. Kaiser and Ulf-G. Mei{\ss}ner,
{\it Phys. Rev. Lett.\/}
{\bf 67} (1991) 1515.
\smallskip
\item{8.}G. Ecker, J. Gasser, A. Pich and E. de Rafael,
{\it Nucl. Phys.\/} {\bf B321} (1989) 311.
\smallskip
\item{9.}J.F. Donoghue, C. Ramirez and G. Valencia,
{\it Phys. Rev.\/} {\bf D39} (1989) 1947.
\smallskip
\item{10.}V. Bernard, N. Kaiser, and Ulf-G. Mei{\ss}ner, {\it Nucl. Phys.\/}
{\bf B373} (1992) 364.
\smallskip
\item{11.}J. Gasser, M.E. Sainio and A. ${\check {\rm S}}$varc,
{\it Nucl. Phys.\/}
{\bf B307} (1988) 779;

Ulf-G. Mei{\ss}ner,
{\it Int. J. Mod. Phys.}
{\bf E1} (1992) 561.
\smallskip
\item{12.}M. Damashek and F. Gilman, {\it Phys. Rev.\/} {\bf D1} (1970) 1319;

T.A. Armstrong et al., {\it Phys. Rev.\/} {\bf D5} (1970) 1640.
\smallskip
\item{13.}
T.A. Armstrong et al., {\it Nucl. Phys.\/} {\bf B41} (1972) 445.
\smallskip
\item{14.}M. N. Butler and M. J. Savage, {\it Phys. Lett.} {\bf B294} (1992)
369.
\smallskip
\item{15.}J.L. Powell, {\it Phys. Rev.} {\bf 75} (1949) 32.
\smallskip
\item{16.}J. Gasser and H. Leutwyler, {\it Ann. Phys. (N.Y.)\/}
 {\bf 158} (1984) 142.
\smallskip
\item{17.}V. Bernard, N. Kaiser and Ulf-G. Mei{\ss}ner,
{\it Phys. Lett.\/} {\bf B309} (1993) 421.
\smallskip
\item{18.}M. Benmerrouche, R.M. Davidson and N.C. Mukhopadhyay,
{\it Phys. Rev.\/} {\bf C39} (1989) 2339.
\smallskip
\item{19.}L.M. Nath and B.K. Bhattacharyya,
{\it Z. Phys.\/} {\bf C5} (1980) 9.
\smallskip
\item{20.}N.C. Mukhopadhyay, A.M. Nathan and L. Zhang,
{\it Phys. Rev.\/} {\bf D47} (1993) R7.
\smallskip
\item{21.}K. Hikasa et al.,
{\it Phys. Rev.\/} {\bf D45} part 2 (1992).
\smallskip
\item{22.}N.M. Butler, M.J. Savage and R. Springer,
{\it Nucl. Phys.\/} {\bf B399} (1993) 69.
\smallskip
\item{23.}V. Bernard, N. Kaiser and Ulf-G. Mei{\ss}ner,
{\it Z. Phys.\/} {\bf C60} (1993) 111.
\smallskip
\item{24.}V. Bernard, N. Kaiser, J. Kambor and Ulf-G. Mei{\ss}ner,
{\it Phys. Rev.\/} {\bf D46} (1992) 2756.
\smallskip
\item{25.}A. Zieger {\it et al.}, {\it Phys. Lett.\/} {\bf B278}
(1992) 34;

F.J. Federspiel {\it et al.}, {\it Phys. Rev. Lett.\/} {\bf 67}
(1991) 1511;

E.L. Hallin {\it et al.}, {\it Phys. Rev.\/} {\bf C48} (1993) 1497.

K.W. Rose {\it et al.}, {\it Phys. Lett.\/} {\bf B234}
(1990) 460;

J. Schmiedmayer {\it et al.}, {\it Phys. Rev. Lett.\/} {\bf 66}
(1991) 1015.
\smallskip
\item{26.}V.A. Petrunkin, {\it Sov. J. Nucl. Phys.\/} {\bf 12} (1981) 278.
\smallskip
\item{27.}R. Erbe et al.,
{\it Phys. Rev.\/} {\bf 188} (1969) 2060.
\bigskip
\bigskip
{\bf FIGURE CAPTIONS} \bigskip
\item{Fig.1}Feynman rules for the $1/m$ suppressed propagators and
vertices. Solid, dashed and wiggly lines refer to nucleons, pions and
photons, in order. Pion momenta are denoted by $q_i$ $(i=1,2,3)$,
nucleon momenta by $\ell$, $p_1$, $p_2$,
and isospin indices by $a,b,c$.
\medskip
\item{Fig.2}The spin--averaged Compton amplitude for the proton
normalized to one at $\omega = 0$. (a) $A(z)$ and $A(\zeta)$ compared
to the data for $0 \le \omega \le 150$ MeV.
(b) $A(z)$ and $A(\zeta)$ compared to the quadratic approximation and
to the data for $70 \le \omega \le 150$ MeV.
\medskip
\item{Fig.3}The spin--averaged Compton amplitude for the neutron.
\medskip
\item{Fig.4}The real part of $A_p (\omega )$
normalized to one at $\omega = 0$ for $\omega > 140$ MeV in comparison
to the data.
\medskip
\item{Fig.5}Dependence of the electromagnetic polarizabilities on
$c_2$ for $\delta \bar{\alpha}_i(m_\Delta)  = \delta \bar{\beta}_i(m_\Delta) =
0$ $(i=p,n)$ in units of $10^{-4}$ fm$^3$.
\medskip
\item{Fig.6}Dependence of the electromagnetic polarizabilities on
$\lambda$ for $\delta \bar{\alpha}_i(\lambda) = \delta \bar{\beta}_i(\lambda)
= 0$
$(i=p,n)$
in units of $10^{-4}$ fm$^3$.
The upper/lower rim of the corresponding bands refers to
$g_A = 1.328 \, / \, 1.26$, in order.
\medskip
\item{Fig.7}
$\delta \bar{\beta}^\Delta ( g_1 , Y )$
in units of $10^{-4}$ fm$^3$
for $3.8 \le g_1 \le 5.3$ and
$-0.8 \le Y \le 0.3$.
\smallskip
\vfill \eject \end